\documentstyle[aps,epsfig,prb]{revtex} 

\newcommand{\extraspace}{\addtolength{\abovedisplayskip}{2mm} 
                        \addtolength{\belowdisplayskip}{2mm} 
                        \addtolength{\abovedisplayshortskip}{2mm} 
                        \addtolength{\belowdisplayshortskip}{2mm}} 
\newcommand{\be}{\begin{equation}\extraspace} 
 
\newcommand{\ee}{\end{equation}} 
 
\newcommand{\bea}{\begin{eqnarray}\extraspace} 
\newcommand{\beastar}{\begin{eqnarray*}\extraspace} 
\newcommand{\eea}{\end{eqnarray}} 
\newcommand{\eeastar}{\end{eqnarray*}} 
 
\newcommand{\nonu}{\nonumber \\[2mm]} 
 
\newcommand{\strutje}{\rule[-1mm]{0mm}{4mm}}

\newcommand{\half}{{\textstyle \frac{1}{2}}} 
\newcommand{\thi}{\frac{1}{3}} 
\newcommand{\quart}{\frac{1}{4}} 
\newcommand{\mm}{\frac{1}{m}} 
\newcommand{\mmm}{\frac{2}{m}}

\newcommand{\eps}{\epsilon}

\newcommand{\gt}{\tilde{g}}
\newcommand{\vac}{| 0 \rangle}

\newcommand{\bra}{\langle}
\newcommand{\ket}{\rangle}

\newcommand{\la}{\lambda}

\newcommand{\del}{\partial}

\begin{document} 
\draft 
\tighten

\title{Quasi-particles in fractional quantum Hall effect edge theories}
 
\author{R. A. J. van Elburg and K. Schoutens} 
 
\address{ Van der Waals-Zeeman Institute and Institute for Theoretical Physics\\
University of Amsterdam, Valckenierstraat 65, 1018 XE
  Amsterdam, The Netherlands} 
\date{\today}

\maketitle

\begin{abstract} 
We propose a quasi-particle formulation of effective edge 
theories for the fractional quantum Hall effect.  
For the edge of a Laughlin state with filling fraction 
$\nu={1 \over m}$,  our fundamental 
quasi-particles are edge electrons of charge $-e$ and edge 
quasi-holes of charge $+{e \over m}$. These quasi-particles 
satisfy exclusion statistics in the sense of Haldane. 
We exploit algebraic properties of edge electrons to 
derive a kinetic equation for charge transport between 
a $\nu={1 \over m}$ fractional quantum Hall edge and a normal 
metal. We also analyze alternative `Boltzmann' equations that 
are directly based on the exclusion statistics properties of edge 
quasi-particles. Generalizations to more general filling 
fractions (Jain series) are briefly discussed.
\end{abstract}

% \vfill
% \noindent ITFA- 
 % \noindent cond-mat/ 
 % \noindent January 1998 
%\narrowtext\twocolumn
%\newpage
\tableofcontents

\newpage 
 
\section{Introduction and summary}
\setcounter{equation}{0}
\setcounter{figure}{0}

Low energy excitations over (fractional) quantum Hall
effect (qHe) ground states are localized near the edge 
of a sample. Certain aspects of qHe phenomenology can 
therefore be captured by an {\em effective edge theory}. 
The unusual properties (notably, fractional charge and 
statistics) of  bulk excitations over a fractional qHe 
(fqHe) ground state carry over to analogous properties of 
the fundamental edge excitations.

In the existing (theoretical) literature on qHe edge 
phenomena, the fundamental edge quasi-particles have 
not played an important role. Most, if not all, results 
have been obtained by exploiting bosonization schemes.
For the analysis of edge-to-edge tunneling experiments,
a combination of bosonization with techniques from
integrable field theories has led to exact results for
universal conductance curves \cite{FLS,dCF}. What has been 
missing until now is a description of edge-to-edge 
transport phenomena directly in terms of excitations 
that are intrinsic to an edge in isolation. The authors of 
\cite{SdCF} have explored such a picture and have derived 
a number of strong-coupling selection rules for the scattering 
of edge excitations.

In this paper we take up the challenge of reformulating
effective qHe edge theories directly in terms of a set of
fundamental edge excitations. For the principal Laughlin 
states at filling fraction $\nu=1/m$, we select the {\em edge 
electron}\ (of charge $-e$) and the {\em edge quasi-hole}\ 
(of charge $+e/m$) as the fundamental excitations. (The 
reason for this asymmetric choice will become clear.) An
important complication is then that these fundamental 
quasi-particles are {\em not}\ fermions but instead obey
fractional statistics (in a sense to be explained below).
A large part of this paper will be devoted to working out 
the consequences of  these unusual statistics.

Following Wen \cite{We1}, we shall assume that the edge theory 
for a $\nu=1/m$ fractional qHe state (Laughlin state) takes the form 
of a chiral Luttinger Liquid (see \cite{SF} for a further justification
of this description). The bosonic description of such a 
theory is centered around the neutral charge density
operators $n_q = {1 \over \sqrt{m}} (\del \varphi)_q$, which 
satisfy a $U(1)$ affine Kac-Moody algebra
\be
 [ n_q , n_{q'} ] = {1 \over m} \, q \, \delta_{q+q'} \ . 
\ee
The vertex operators
\bea
&&\Psi_{\rm e}(z) = \, : \exp(-i\sqrt{m}\varphi) :(z) \ , \qquad 
\Psi_{\rm qh}(z) = \, : \exp(i{1 \over \sqrt{m}}\varphi) :(z)
\label{qp}
\eea
have charge $-e$ and $+{e \over m}$, respectively, and have 
been identified as the edge electron and edge quasi-hole.
They are the direct analogues of the bulk (Laughlin)
quasi-particles, an important distinction being that
the edge excitations are gap-less and have linear dispersion.

The first indication for the non-trivial statistics of the
operators (\ref{qp}) comes from the Operator Product Expansions 
(OPE)
\bea
&&
\Psi_{\rm e}(z) \Psi_{\rm e}(w) = (z-w)^m 
  \, [ \Psi'(w) + \ldots ]\ , \qquad
\Psi_{\rm qh}(z) \Psi_{\rm qh}(w) = (z-w)^{1 \over m} 
  \, [ \Psi''(w) + \ldots ]
\label{exc}
\eea
where $\Psi'(w)$ and $\Psi''(w)$ are operators of charge 
$-2e$ and $+{2e \over m}$, respectively.
For $m$ an odd integer, the right hand side of the first OPE 
picks up a minus sign under the exchange $z \leftrightarrow w$, 
in correspondence with the required antisymmetry of the Laughlin 
wave functions. The second relation features a fractional power 
of $(z-w)$, which shows that the quasi-hole operator has 
fractional `exchange statistics'.

For $\nu=1$ the operators $\Psi_{\rm e}$ and $\Psi_{\rm qh}$ are fermionic
and the edge theory is simply a theory of free fermions. The
exchange statistics of $\Psi_{\rm e}$ and $\Psi_{\rm qh}$ at $\nu=1/m$ 
clearly signal a deviation from free fermion behavior, but they are 
hardly helpful for the purpose of setting up a quasi-particle 
formalism that mimics the free fermion treatment of the $\nu=1$ 
edge. We shall here argue that a much more convenient point of 
view is that of the `exclusion statistics' properties of $\Psi_{\rm e}$ 
and $\Psi_{\rm qh}$.

In recent paper \cite{Sc}, one of us proposed a method to 
associate exclusion statistics to quasi-particles for Conformal
Field Theory (CFT) spectra. We shall here show that, 
when applied to the $\nu=1/m$ edge theory excitations, this  
method gives ideal fractional exclusion statistics (in the sense of 
Haldane \cite{Ha1}) with $g=m$ for edge electrons and 
$g={1 \over m}$ for edge quasi-holes. 
We also find that the edge electrons 
and edge quasi-holes can be viewed as independent excitations, 
in the sense that there is no mutual exclusion between the two.  

Our program in this paper is then (1) to establish the exclusion 
statistics properties of $\Psi_{\rm e}$ and $\Psi_{\rm qh}$, and 
(2) to apply them to both equilibrium and transport properties of 
these edges. As for transport, we shall focus on the set-up
of the experiment by Chang et al. \cite{CPW}, where electrons 
are allowed to tunnel from a normal metal into the edge of a 
$\nu=1/3$ fqHe edge. We shall use algebraic properties of $\nu=1/3$ 
edge electrons to write an exact kinetic equation for the perturbative
$I$-$V$ characteristics for this system, reproducing the result 
obtained by other methods. Interestingly, the relevant algebraic 
properties derive from the so-called $N=2$ superconformal algebra, 
which has been well-studied in the context of String Theory. 
We shall also study `naive' Boltzmann equations
that are based on the exclusion statistics properties of the
edge quasi-particles. While the latter equations are not exact, we 
shall argue that they can be used as the starting point in a 
systematic approximation to the exact transport results. These
results then are of general importance, as they illustrate
the possibilities and limitations of the concept of
exclusion statistics in the analysis of non-equilibrium physics.

The observations made here are easily generalized to composite edges,  
related to hierarchical fqHe states, in particular those of the Jain 
series with $\nu={n \over np+1}$. For the Jain series edge theories, 
two natural pictures emerge. In the first picture, the edge 
quasi-particles satisfy Haldane's exclusion statistics 
with $G$-matrix equal to $K^{-1}$, where $K$ is the topological order 
matrix of the bulk fqHe state. In the second picture one 
decouples one charged mode from $n-1$ neutral modes. 
A possible quasi-particle basis then consists of a single charged 
Haldane $g$-on and a collection of $n$ neutral quasi-particles that
are related to parafermions in the sense of Gentile.

This paper is organized in the following way. 
In section 2 we discuss exclusion statistics and indicate the 
applications to fqHe states and to CFT spectra. In section~3  
we discuss in some detail how a quasi-particle basis for 
$\nu={1 \over m}$ fqHe edge states is obtained and how that leads 
to an assignment of exclusion statistics parameters. 
In section~4 we explain how equilibrium properties are obtained
in a quasi-particle approach. In section~5 we further study the 
quasi-particle bases and make the link with Calogero-Sutherland 
quantum mechanics and Jack polynomials.
In section 6 we study charge transport between a normal metal and a 
$\nu={1 \over m}$ fqHe edge in terms of kinetic equations that 
are based on our quasi-particle formalism.  The appendix A 
describes the extension of our quasi-particle formalism to filling 
fractions in the Jain series, while appendix B contains 
explicit results for an important quasi-particle form factor.

%\newpage

\section{Exclusion statistics} 
\setcounter{equation}{0}
\setcounter{figure}{0}

In his by now famous 1991 paper \cite{Ha1}, Haldane
proposed the notion `fractional exclusion statistics',
as a tool for the analysis of strongly correlated 
many-body systems. The central assumption that
is made concerns the way a many-body spectrum is
built by filling available one-particle states. In words, it
is assumed that the act of filling a one-particle state 
effectively reduces the dimension of the space of remaining 
one-particle states by an amount $g$. The choices $g=1$, $g=0$ 
correspond to fermions and bosons, respectively. The 
thermodynamics for general `$g$-ons', and in particular the
appropriate generalization of the Fermi-Dirac distribution
function, have been obtained in \cite{Wu,NW,Isa,Ra}. The so-called 
Wu equations \cite{Wu}
\bea
&&n_g(\eps) = {1 \over [w(\eps)+g]}\qquad {\rm with} \qquad [w(\eps)]^g[1+w(\eps)]^{1-g} = e^{\beta(\eps-\mu)}
\label{wueq}
\eea
provide an implicit expression for the 1-particle distribution
function $n_g(\eps)$ for $g$-ons at temperature $T$ and
chemical potential $\mu$. It has been demonstrated that fractional 
exclusion statistics are realized in various models for Quantum 
Mechanics with inverse square exchange \cite{Ha1,Ha,MS} and in the 
anyon model in a strong magnetic field \cite{DO,Wu}.

\subsection{Exclusion statistics and the fqHe}
 
A natural application of the idea of exclusion statistics
is offered by the various fractional quantum Hall effects. 
One may take the somewhat naive but certainly justifiable 
point of view that the essence of the $\nu=\mm$ fqHe is that
under the appropriate conditions interacting electrons
give rise to free quasi-particles with effective statistics
parameter $g=m$. A familiar interpretation of these 
quasi-particles is that they can be viewed as composites of 
electrons plus an even number of flux quanta \cite{ftn1}.
The familiar Laughlin wave functions describe the ground state 
configuration for these quasi-particles. The fundamental 
excitations (the Laughlin quasi-particles) are expected to 
carry the `dual' (see section 3.3) statistics $g=\mm$.

The above scenario, which was suggested in Haldane's
original paper, has been critically analyzed in the 
literature, where it has been confirmed  in the
appropriate low-temperature regime (see for example
\cite{statsfqhe}). Our purpose in 
this paper is to set up and analyze a similar picture 
for {\em edge excitations}\ in the fqHe. Since such
excitations can be described using the language of 
CFT, we first turn to a discussion of exclusion
statistics in CFT spectra.

\subsection{Exclusion statistics in CFT}

Conformal Field Theories in two dimensions come with two
commuting Virasoro algebras, and these infinite dimensional
algebras can be used to organize the finite-size spectra of these 
theories. In such an approach, a CFT partition function is obtained 
by combining a number of characters of both Virasoro algebras
(or extensions thereof). In applications such as String Theory,
where the conformal symmetry has a geometric origin and the
fundamental fields are bosonic coordinate fields, this
`Virasoro approach' to CFT is entirely natural. In contrast, the
prototypical CFT in the condensed matter arena is a theory 
of free fermions, with a finite size spectrum that is simply a
collection of many-fermion states constructed according to
the rules set by the Pauli principle. When facing other CFT's that 
are relevant for condensed matter systems one may try to follow
a similar road, which is to select a number of fundamental
quasi-particle operators and to construct the full (chiral) 
spectrum as a collection of many-(quasi)-particle states. Explicit
examples of this are the so-called spinon bases for a 
$\widehat{su(n)}_k$ Wess-Zumino-Witten models 
\cite{Ha2,BPS,BLS,BS}. 

Let us now imagine that we have a concrete CFT, with explicit
rules for the construction of a many-(quasi-)particle basis
of the finite size spectrum. It is then natural to try to
interpret that result in terms `exclusion statistics' 
properties of the fundamental quasi-particles. In a recent
paper \cite{Sc}, one of us has proposed a systematic procedure 
(based on recursion relations for truncated chiral spectra), 
which leads to 1-particle distribution functions for CFT 
quasi-particles. In many cases, it was established that 
the CFT thermodynamics are those of a free gas of quasi-particles 
governed by the new, generalized distribution functions.
The examples discussed in \cite{Sc} include spinons in the 
$\widehat{su(n)}_1$ WZW models, CFT parafermions and 
edge quasi-particles for the fractional qHe.

The example of the CFT for fqHe edge excitations is particularly 
interesting, since in those cases the generalized distributions 
derived from the CFT spectra are identical to those obtained from 
Haldane statistics (with specific values for $g$). In 
section 3 below we show in some detail how these results 
are established. 

Clearly, the identification of Haldane statistics in fqHe edge
theories is most useful since it provides a concrete link between 
rather abstract considerations on the systematics of quasi-particle 
bases on the one hand and concrete laboratory physics on the other. 
In particular, it opens up the possibility of analyzing transport
phenomena such as edge-to-edge tunneling in the qHe (which has been 
well-studied both theoretically and experimentally) directly in 
terms of quasi-particles satisfying fractional exclusion statistics. 
We shall report the results of such an analysis in section 6 below.

%\newpage

\section{Quasi-particles for the $\nu={1 \over m}$ fqHe edge}
\setcounter{equation}{0}
\setcounter{figure}{0}

We consider the finite size spectrum for the CFT describing a
single $\nu={1 \over m}$ fqHe edge. In the CFT jargon, this theory
is characterized as a $c=1$ chiral free boson theory at radius 
$R^2=m$. We shall consider the chiral Hilbert space corresponding
to the following partition function 
\be
Z^{1/m}(q) = \sum_{Q=-\infty}^{\infty} {q^{Q^2 \over 2m}
                 \over (q)_{\infty}} \ , 
\label{Zm}
\ee
with $(q)_\infty = \prod_{l=1}^{\infty}(1-q^l)$ and 
$q=e^{-\beta {2\pi \over L}{1 \over \rho_0}}$. [The 1-particle
energies are of the form $\eps_l=l {2\pi \over L}{1 \over \rho_0}$
with $l$ integer and $\rho_0$ the density of states per unit length,
$\rho_0=(\hbar v_F)^{-1}$.] In this formula, the $U(1)$ affine 
Kac-Moody symmetry is clearly visible as all states at fixed 
$U(1)$ charge $Q$ form an irreducible representation of this 
symmetry. 

We should stress that the Hilbert space corresponding to (\ref{Zm}) 
is not the physical Hilbert space for the edge theory of a quantum 
Hall sample with the topology of a disc. In the latter Hilbert space 
physical charge is quantized in units of $e$ and, correspondingly, 
the $U(1)$ charge $Q$ in (\ref{Zm}) is restricted to multiples of 
$m$ \cite{We1}. In the geometry of a Corbino disc, {\em i.e.}, a cylinder, 
the operator that transfers charge ${e \over m}$ from one edge to 
the other is physical. Accordingly, the physical Hilbert space is 
obtained by taking a tensor product of left and right copies of the 
Hilbert space (\ref{Zm}) and restricting the total $U(1)$ charge 
$Q_L+Q_R$ to multiples of $m$ \cite{We1}. In the quasi-particle
formalism that we present below the various restrictions on $Q_L$, 
$Q_R$ are easily implemented.

Our goal here is to understand the collection of states in 
(\ref{Zm}) in a different manner, and to view them as multi-particle 
states built from the creation operators for edge quasi-particles 
$\Psi_{\rm e}$ and $\Psi_{\rm qh}$. To simplify our notations, 
we shall write $G \equiv \Psi_{\rm e}$, $\phi \equiv \Psi_{\rm qh}$
\cite{ftn2}.
Due to the above-mentioned
restrictions on the $U(1)$ charges $Q$ and $Q_L+Q_R$, the chiral 
quasi-hole operator $\phi(z)$ by itself is not a physical operator 
in the edge theories for the disc or cylinder \cite{ftn3}
physical states are obtained by restricting the number of 
$\phi$-quanta in the appropriate manner.

\subsection{Quasi-hole states}

We start by considering quasi-hole states that are built by 
applying only the modes $\phi_{-s}$ defined via $\phi(z) 
= \sum_s \phi_{-s} z^{s-{1 \over 2m}}$. 
Clearly, the index $s$ gives the dimensionless 
energy of the mode $\phi_{-s}$. When acting on the charge-$0$ 
vacuum $|0\rangle$, we find the following multi-$\phi$ states 
(compare with \cite{BLS} for the case $m=2$, see also \cite{Is})
\bea
&&
 \phi_{-{(2N-1) \over 2m}- n_N} \ldots
 \phi_{-{3 \over 2m}- n_2} 
 \phi_{-{1 \over 2m}- n_1} | \, 0 \, \rangle
\  \qquad {\rm with}\ \quad 
   n_N \geq n_{N-1} \geq \ldots \geq n_1 \geq 0 \ . 
\label{phistates}
\eea
The choice of minimal modes is such that the lowest state of  charge 
$Q{e \over m}$ is at energy ${Q^2 \over 2m}$, in agreement with the 
scaling dimension of the corresponding CFT primary field. Using so-called
generalized commutation relations satisfied by the modes $\phi_{-s}$
one may show \cite{BLS} that all multi-$\phi$ states different from 
(\ref{phistates}) are either zero or linearly dependent on 
(\ref{phistates}).

Before writing more general states we shall first focus on the
exclusion statistics properties of the quanta $\phi_{-s}$. We 
follow the procedure of \cite{Sc} and start by introducing truncated 
partition sums for quasi-hole states (\ref{phistates}).
For $s={1 \over 2m}$, ${3 \over 2m}$, etc, we define polynomials 
$P_s(x,q)$ to keep track of the number of many-body states that 
can be made using only the modes $\phi_{-k}$
with $k \leq s$, and that have a highest occupied mode with energy
$s'$ such that $s-s'$ is integer. $P_s(x,q)$ is defined as the 
trace of the quantity $x^N q^E$ over all these states, where $N$ is 
the number of quasi-holes, $E$ is the dimensionless total energy,
and  $x=e^{\beta\mu_{\rm qh}}$. For $m=3$ this gives 
\be
 P_{{1\over 6}}= xq^{1\over 6}\  ,
\quad
 P_{{1\over 2 }}= x^2 q^{4\over 6}\ ,
\quad 
 P_{{5\over 6}}= 1+x^3 q^{9\over 6} \ , \quad {\rm etc}.
\ee
In general, a occupied quasi-hole state of energy $s$ corresponds to a factor $xq^s$ in these generating polynomials.

The systematics of the edge quasi-hole states (\ref{phistates})
directly lead to the following
recursion relations between the polynomials $P_s(x,q)$, 
\bea 
  P_s(x,q)=P_{s-1}(x,q)+ x \, q^s \, P_{s-{1 \over m}}(x,q) \  .
\eea
For $m=1$, which is the case corresponding to a $\nu=1$ 
integer qHe edge, this relation directly implies
$P_{l-\half}(x,q) = \prod_{j=1}^l (1+x q^{j-\half})$.
In that case the partition sum is simply a product and
we recognize free fermions. For general $m$ things are not that
simple, but we can rewrite the recursion relation in 
matrix form
\be
   \pmatrix{ P_{l-{2m-1 \over 2m}}\cr \vdots\cr P_{l-{1 \over 2m}}} 
    =M^{\rm qh}_l(x,q)
   \pmatrix{P_{l-1-{2m-1 \over 2m}}\cr \vdots  \cr P_{l-1-{1 \over 2m}}},
\ee
with $l=1,2,\ldots$ and $M_l^{\rm qh}(x,q)$ the following $m\times m$ matrix
%\widetext
\bea
  \lefteqn{M^{\rm qh}_l(x,q)=}
\nonu &&
  \pmatrix{
   \strutje 1&    0&    \ldots&    0&    x q^{l-{2m-1\over 2m}}\cr 
   \strutje xq^{l-{2m-3\over 2m}}&   1&\ldots& 0&  
   x^2q^{2l-{4m-4\over 2m}} \cr  \cr
   \vdots& &\ddots& &\vdots\cr \cr
   x^{m-1}q^{(m-1)l-{m\over 2}+{2m-1 \over 2m}} &
   x^{m-2}q^{(m-2)l-{m \over 2}+{4m-4 \over 2m}}&\ldots&
   x q^{l-{1 \over 2m}} &
   1+ x^m q^{ml-{m \over 2}} } .
\nonu
&&
\eea
%\narrowtext
The grand partition function for the quasi-hole states (\ref{phistates}) 
is then given by
\be
  Z^{\rm qh}(x,q)=\pmatrix{1&1&\ldots&1}
                    \left(\prod_{l=1}^\infty
                    M_l^{\rm qh}(x,q)\right)
                    \pmatrix{0\cr  \vdots\cr0 \cr 1}.
\label{Zqh}
\ee   
We propose that the quasi-hole modes $\phi_{-s}$ with $s=l-{2m-1 \over 2m},
\ldots, l-{1 \over 2m}$ be viewed as a single ($m$-fold degenerate) level 
in the one-particle spectrum. [This convention is natural since a single
quasi-particle over the ground state can only occupy one of these
$m$ levels.] The $m\times m$ matrix $M_l^{\rm qh}$
is then a level-to-level transfer matrix and replaces the free fermion
($m=1$) factor $(1+xq^{l-\half})$. Clearly, the thermodynamics of
the states (\ref{phistates}) will be dominated by the largest
eigenvalues $\lambda_l^+(x,q)$ of the matrices $M_l^{\rm qh}(x,q)$.
These satisfy the characteristic equations
\be
  (\lambda_l^+ -1)^m - x^m q^{ml-{m\over 2}} (\lambda_l^+)^{m-1}=0 \ .
\ee
Instead of trying to solve these equations, we can derive from
them a result for the 1-particle distribution 
functions 
\be
  n^{\rm qh}(l) \equiv x \partial_x \ln (\lambda_l^+)
                = x {\partial_x \lambda_l^+ \over \lambda_l^+} \ .
\label{xdelx}
\ee
We find
\bea
&&  n^{\rm qh}(l) = {\lambda_l^+-1 \over 1+ {1\over m}(\lambda_l^+-1 )},
  \nonu 
&&  (x q^l)^{-1} = (\lambda_l^+-1)^{-1} 
                        (\lambda_l^+)^{1-{1 \over m}} \ .
\eea
Comparing with (\ref{wueq}) and identifying $g={1 \over m}$ and
$w(\eps) = (\lambda_l^+-1)^{-1}$, we see that the distribution 
function $n^{\rm qh}(l)$ becomes identical to 
$n_{g={1 \over m}}(\eps=l)$. In other words, the exclusion statistics 
properties of the 
$\nu={1 \over m}$ quasi-holes are those of `ideal $g$-ons'
in the sense of Haldane, with $g={1 \over m}$! This identification is
consistent with the result of bosonization applied to $g$-ons 
\cite{WY}, and with the character computations of \cite{Hi}.

For the case $m=2$, which is not in the category of fqHe edges,
the equilibrium distribution is given by
\be
n_{\half}(\eps) = {2  \over  \sqrt{1+4e^{-2\beta(\mu-\eps)}}} \ .
\ee
For $m=3$ the explicit formulas (obtained using the Cardano
formula for cubic equations) are quite unpleasant; figure~3.1 
shows the distribution $n_{\thi}(\eps)$.

\begin{figure}[ht]
\setlength{\unitlength}{1pt}
\centerline{
\begin{picture}(450,250)
\put(0,-40){\psfig{file=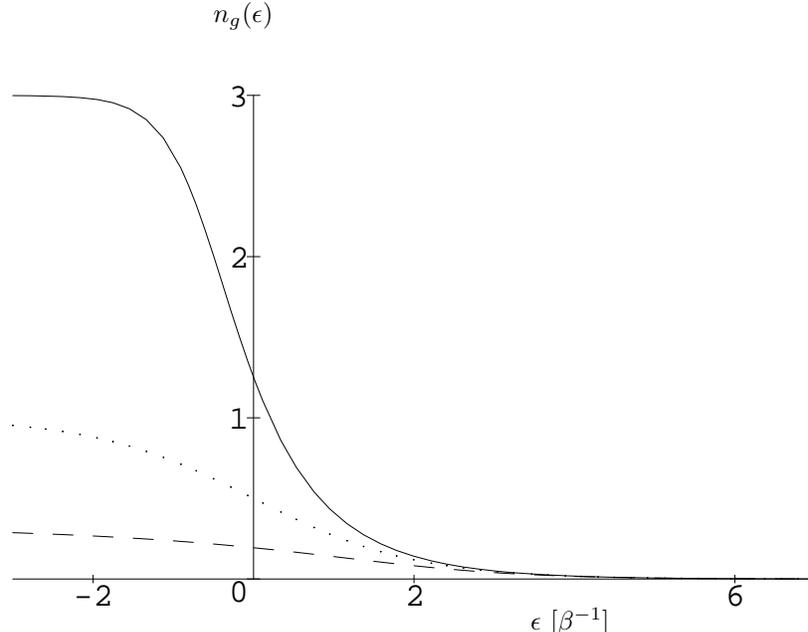}}
\put(270,0){$\epsilon \ [\beta^{-1}]$}
\put(150,230){{$n_g(\epsilon)$}}
\end{picture}}
\caption{Distribution functions for fractional exclusion statistics 
with $g=3$ (dashed line), $g=1$ (dotted line), and $g={1 \over 3}$
(solid line) , all at the same temperature and at zero chemical 
potential.}
\end{figure}

\subsection{Edge electron states}

The same procedure can be applied to the edge electrons,
which are created by modes $G_{-t}$ with 
$G(z)=\sum_t G_{-t} z^{t-{m \over 2}}$. Multi-electron states 
take the form
\bea
&&
 G_{-(2M-1){m \over 2} - m_M} \ldots
 G_{-3{m \over 2}- m_2} 
 G_{-{m \over 2} -m_1} | \, 0 \, \rangle
 \qquad {\rm with}\ \quad 
   m_M \geq m_{M-1} \geq \ldots \geq m_1 \geq 0 
\eea
and we have truncated partition sums $Q_t(y,q)$ with
$t$  a half-odd-integer and $y=e^{\beta \mu_{\rm e}}$. They
satisfy the recursion relations
\bea 
  Q_t(y,q)=Q_{t-1}(y,q)+ y \, q^t \, Q_{t-m}(y,q),
\eea
with the following initial values 
\be
Q_{-{m\over 2}}= \ldots =Q_{{m\over 2}-1}=1 \ .
\ee
The 'transfer matrix' for the edge electrons 
$M^{\rm e}_k(y,q)$ is defined by
\be
   \pmatrix{Q_{K}\cr\vdots\cr Q_{K+m-1}} 
   = M^{\rm e}_k(y,q)
   \pmatrix{Q_{K-m}\cr\vdots\cr Q_{K-1}} 
\ee
with $K={km - {m/2}}$, $k=1,2,\ldots$
and we have
\be
  Z^{\rm e}(y,q)=\pmatrix{1&0 \ldots&0&0}
           \left(\prod_{k=1}^\infty M^{\rm e}_k(y,q)\right)
           \pmatrix{1\cr  \vdots\cr 1\cr 1} .
\label{Ze}
\ee
In this case, a single action of the transfer matrix comprises 
a jump of $m$ 1-particle levels, and the relevant distribution 
function will be 
\be
  n^{\rm e}(k) \equiv y \partial_y \ln [ (\mu_k^+)^{1 \over m} ]
               = {y\over m} {\partial_y \mu_k^+ \over \mu_k^+} ,
\ee
The characteristic equation for the eigenvalue $\mu_k^+$
\be
  \prod_{i=0}^{m-1}(\mu_k^+ - y q^{mk-i}) - (\mu_k^+)^{m-1} =0 \ ,
\ee
leads to 
\be
  n^{\rm e}(k) = {1 \over m + (h_k-1)} \ ,
  \quad 
  (y q^{mk})^{-1} = (h_k -1)^m h_k^{1-m}.  
\ee
with $h_k= \mu_k^+/(y q^{mk})$. Identifying 
$w(\eps)= h_k-1$, we again recognize the Wu equations (\ref{wueq}) 
for Haldane exclusion statistics, this time with $g=m$, and we 
may identify $n^{\rm e}(k)$ with $n_{g=m}(\eps=mk)$.

For $m=2$ this gives
\be
n_2(\eps) = \half \left( 
   1 - {1 \over \sqrt{ 1+4\, e^{-\beta(\eps-\mu)}}} \right) \ .
\ee 
See figure~3.1 for the distribution 
function $n_3(\eps)$ at $\mu_{\rm e}=0$.

\subsection{Duality}

Having recognized distribution functions for fractional
exclusion statistics with $g={1 \over m}$ and $g=m$, 
respectively, we expect a particle-hole duality between 
the two cases (compare with \cite{NW,Ra}). 

Before we come to that, we generalize the results of 
sections 3.1 and 3.2 by considering a chiral
$c_{CFT}=1$ CFT of compactification radius $R^2=r/s$,
with $r>s$ and $r,s$ coprime. Choosing $\phi$-quanta
of charge $+{s \over r}e$ and $G$-quanta of charge $-e$ 
as our fundamental excitations, we easily repeat
the previous analysis and derive the following 
recursion relations
\be
X_l(x) = X_{l-r}(x) + x \, X_{l-s}(x)\ , \qquad
Y_l(y) = Y_{l-s}(y) + y \, Y_{l-r}(y) \ ,
\label{recurs}
\ee
where we put $q=1$ for convenience. [The connection with
the quantities $P_l$ and $Q_l$ defined for ${r \over s}=m$
is $X_l \leftrightarrow P_{{2l-1 \over 2m}}$,
$Y_l \leftrightarrow Q_{{m \over 2}+(l-1)}$.]
Proceeding as before we obtain the distribution functions
for Haldane statistics with $g=s/r$ (for the $\phi$-quanta)
and $\gt=r/s$ (for the $G$-quanta).

In the papers \cite{NW,Ra} it was recognized that the cases 
with $g$ and $\gt=1/g$ are dual in the sense that particles 
are dual to holes.
To recover this duality in our present approach, we
note that if $Y_l(y)$ is a solution of the second relation
in (\ref{recurs}), the expression
\be
X_l(x) = Y_l(y=x^{-\gt})  \, x^{l \over s}
\ee
solves the first recursion relation. Assuming $r>s$,
we can rewrite both recursion relations in a form
involving a $r\times r$ recursion matrix. The largest
eigenvalues $\lambda^+(x)$ and $\mu^+(y)$
are then related via
\be
\lambda^+(x) = \mu^+(x^{-\gt}) \, x^{\gt}
\ee
and the distribution functions
\be
n_g(x) \equiv x\partial_x \ln \lambda^+(x) \ ,
\qquad
n_{\gt}(y) \equiv g \, y\partial_y \ln \mu^+(y) \ ,
\ee
satisfy
\be
g \, n_g(x) = 1 - \gt \, n_{\gt}(y=x^{-\gt}) \ ,
\ee
or, putting $\mu_{\gt}=-\gt \mu_g$ and restoring
$q \neq 1$
\be
g \, n_g(\eps) = 1 - \gt \, n_{\gt}(-\gt \eps) \ ,
\label{dual}
\ee
in agreement with the results of \cite{NW}. The interpretation 
of this result is that the $\gt$ quanta with positive 
energy act as holes in the ground state distribution of negative 
energy $g$-quanta. The relative factor $(-\gt)$ between the energy
arguments in (\ref{dual}) indicates that the act
of taking out $r$ $g$-quanta corresponds to
adding $s$ $\gt$-quanta. This duality further implies
that, when setting up a quasi-particle description for
fractional qHe edges, we can opt for (i) either quasi-holes
\`or edge electrons, with energies over the full range
$-\infty < \eps < \infty$, or (ii) a combination of both
types of quasi-particles, each having positive energies 
only. The CFT finite size spectrum naturally leads to
(ii) (see section 3.4 below), while the analogy with
Calogero-Sutherland quantum mechanics naturally leads
to option (i) (see section 5). When considering transport 
equations in section 6, we shall be considering both 
alternatives.

\subsection{The full spectrum}

To complete our quasi-particle description for the
$\nu={1 \over m}$ edge, we need to specify 
how quasi-hole and electron operators can be combined to
produce a complete basis for the chiral Hilbert space
(\ref{Zm}). We consider the following set of states
%\widetext
\bea
&& 
 G_{-(2M-1){m \over 2}+Q -m_M} \ldots
           % G_{-3{m \over 2}-Q -m_2} 
 G_{-{m \over 2}+Q -m_1} 
 \phi_{-(2N-1){1 \over 2m}-{Q \over m}-n_N} \ldots
           % \phi_{-3 {1 \over 2m}-{Q \over m}-n_2} 
 \phi_{-{1 \over 2m}-{Q \over m}-n_1} | \, Q \, \rangle
\nonu
&& \quad {\rm with}\ \quad 
   m_M \geq m_{M-1} \geq \ldots \geq m_1 \geq 0 ,\quad
   n_N \geq n_{N-1} \geq \ldots \geq n_1 \geq 0  \ ,
\nonu
&& \qquad \qquad  n_1 >0 \quad {\rm if}\ \ Q < 0 \ ,
\label{listofstates}
\eea
%\narrowtext
where $|Q\rangle$ denotes the lowest energy state of charge
$Q {e \over m}$ with $Q$ taking the values $-(m-1)$,
$-(m-2)$, $\ldots$, $-1$, $0$. Our claim 
is now that the collection (\ref{listofstates}) forms a basis
of the chiral Hilbert space, so that
\bea
  Z^{1/m}(q)= \sum_{Q=-(m-1)}^{0} \, q^{Q^2\over 2m} \,
           Z^{\rm qh}_Q(x=1,q) \, Z^{\rm e}_Q(y=1,q) \nonu
\label{Zid}
\eea
where we added a factor $q^{Q^2\over 2m}$ to take into account the
energy of the initial states and we denoted by $Z^{\rm qh}_Q$
and $Z^{\rm e}_Q$ the generalizations of the partition functions
(\ref{Zqh}) and (\ref{Ze}) to the sector with vacuum charge $Q$.
They are naturally written as
\bea
&&Z^{\rm qh}_Q = \sum_{N=0}^\infty
{q^{ {1 \over 2m}(N^2+2QN)+(1-\delta_{Q,0})N } \over (q)_N}, \nonu 
&&Z^{\rm e}_Q = \sum_{M=0}^\infty
{q^{ {m \over 2} M^2 -QM }  \over (q)_M} \ ,
\eea
with $(q)_L=\prod_{l=1}^L (1-q^l)$.

While the collection of states (\ref{listofstates}) looks rather complicated, 
it may be understood by considering the special case $m=1$, which 
is a theory of two real free fermions of charge $\pm 1$. In this case 
there is only the $Q=0$ vacuum and the allowed $\phi$ and $G$ modes
reduce to the familiar free fermion modes $\psi^{\pm}_{-\half-n_j}$.

The right hand side of (\ref{Zid}) has the form of a
so-called `fermionic sum formula' \cite{KKMM} and the equality of
(\ref{Zm}) and (\ref{Zid}) is a new Rogers-Ramanujan identity. 
Similar identities relating `fermionic sums' to characters in 
conformal field theories have been studied in the literature,
see for example \cite{KKMM,BLS,BLS2}. We would like to stress that the
reasoning leading to these identities is very different between our
approach and the work of \cite{KKMM}: in our approach the identities
express exclusion statistics properties of CFT fields, while in the
work of Kedem et al. the identities are based on Bethe Ansatz
solutions of specific integrable lattice models. The first example
where these two approaches have been explicitly connected is that of
spinons in $SU(2)_1$ CFT and in the associated Haldane-Shastry spin 
chains \cite{Ha2,BPS,BLS}.

The important conclusion from the above is that, up to a finite
sum over vacuum charges, the chiral partition sum factorizes
as a product of a quasi-hole piece and an edge electron piece. This
means that the two types of quasi-particles are independent,
or, in other words, that they do not have any mutual exclusion statistics.
This then explains our asymmetric choice of quasi-particles.
Had we chosen to work with fundamental quasi-particles of
charges $\pm {e \over m}$, we would have come across non-trivial
mutual statistics. All of this is nicely illustrated with
the case $m=2$ where we can opt for the `qHe basis' with
independent quasi-holes and edge electrons, or for a
`spinon basis' built from charge $\pm {e \over 2}$ quanta, 
which are identical to the spinons of \cite{Ha2,BPS,BLS} and which
have a non-trivial $2 \times 2$ statistics matrix. The two 
choices have the quasi-hole states (\ref{phistates}) 
(called `fully polarized spinon states' in \cite{BLS}) 
in common, but differ in the way negative charges are brought 
in.

The observations made in this section may be generalized to 
composite edges such as those of the so-called 
Jain series with filling fraction $\nu={n \over np+1}$. 
We refer to appendix~A for a brief discussion.

%\newpage

\section{Equilibrium quantities}
\setcounter{equation}{0}
\setcounter{figure}{0}

\subsection{Specific heat}

The specific heat of a conformal field theory is well-known to 
be proportional to the central charge $c_{CFT}$
\be
{C(T) \over L} 
= \gamma \rho_0 k_B^2 T \ , \quad
\gamma = {\pi \over 6} \, c_{CFT} \ ,
\ee
where $\rho_0=(\hbar v_F)^{-1}$ is the density of
states per unit length.

In \cite{IAMP} it was shown that the specific heat for $g$-on
excitations, with energies in the full range $-\infty
< \eps < \infty$, is in agreement with the central charge 
$c_{CFT}=1$ of the corresponding CFT. The same result should of 
course come out in a picture where we select positive energy 
electrons and positive energy quasi-holes as our fundamental 
excitations. In this picture the total energy carried by the 
edge quasi-particles takes the form
\be
E = \rho_0 \int_0^\infty d\eps \, \eps \, n_g(\eps)
    +
      \rho_0 \int_0^\infty d\eps \, \eps \, n_{\gt}(\eps) \ .
\ee
and the corresponding result for the specific heat is
\be
{C(T)\over L } = (\gamma_{g,+} + \gamma_{\gt,+}) \rho_0 k_B^2 T \ ,
\ee
where 
\be
\gamma_{g,+} = \del_\beta \int_0^\infty 
  d\eps \, \eps \, n_g(\eps)
\ , \qquad \gamma_{\gt,+} = \del_\beta \int_0^\infty
  d\eps \, \eps \, n_{\gt}(\eps) \ .
\ee
It takes an elementary application of the duality relation
(\ref{dual}) to show that $\gamma_{\gt,+}=\gamma_{g,-}$
and hence
\be
(\gamma_{g,+} + \gamma_{\gt,+}) = \gamma_g = {\pi \over 6} \ ,
\ee
confirming once again the value $c_{CFT}=1$. 

We would like to stress that the individual contributions
$\gamma_{g,+}$ do depend on $g$ and that only for $g=1$
(Majorana fermions) $\gamma_{g,+}$ and $\gamma_{g,-}$
are equal. An exact result is \cite{NRT}
\be
\gamma_{g,+} = {\pi \over 6} {L(\xi^g) \over L(1)} \ ,
\ee
with $\xi$ a solution of the algebraic equation
\be
\xi^g = 1-\xi
\ee
and $L(z)$ the Rogers dilogarithm. This gives
\be
\gamma_{\half,+} = {\pi \over 6} \, {3 \over 5} ,\quad
\gamma_{2,+} = {\pi \over 6} \, {2 \over 5} , \quad
\gamma_{\thi,+} = {\pi \over 6} \, 0.655\ldots , \quad
\gamma_{3,+} = {\pi \over 6} \,  0.344\ldots , \quad {\rm etc.}
\ee

\subsection{Hall conductance}

While the specific heat coefficient $\gamma$ is not
sensitive to $g$, the edge capacitance or, equivalently, 
the Hall conductance, obviously does depend on the filling 
fraction $\nu$ and thereby on $g$. In the quasi-particle
formulation, this result comes out in a particularly 
elegant and simple manner.

Let us focus on a $\nu={1 \over m}$ edge and take as our
fundamental quasi-particles the edge electron of charge
$q=-e$ and statistics $g=m$ and the edge quasi-hole
of charge $q={e \over m}$ and statistics $\gt={1 \over m}$,
all quasi-particles having positive energies only.

Let us first consider zero temperature, where the Haldane
distribution functions are step-functions with maximal
value $n_g= {1 \over g}$. If we now put a voltage
$V>0$ the $q<0$ quasi-particles will see their Fermi energy
shift by the amount $qV$ and all available states
at energy up to $-qV$ will be filled. The total charge
$\Delta Q(V,T=0)$ that is carried by these 
excitations equals \cite{ftn4}
\be
\Delta Q(V,T=0) = {1 \over g} \cdot q \cdot \rho_0 \cdot (-qV)
\ee
where the factor ${1 \over g}$ originates from the maximum
of the distribution function and thus represents the 
statistics properties of the quasi-particles. Clearly,
positive-$q$ quasi-particles do not contribute to
the response at $T=0$, $V>0$.

For the $\nu={1 \over m}$ fqHe edges, the result for
$V>0$ is 
\be
\Delta Q(V>0,T=0) = {1 \over m} \cdot (-e) \cdot \rho_0 \cdot (eV)
= - {e^2 \over m} \rho_0 V
\ee
while for $V<0$
\be
\Delta Q(V<0,T=0) = m \cdot {e \over m} \cdot \rho_0 \cdot (- {e \over m} V)
= - {e^2 \over m} \rho_0 V \ .
\ee
Clearly, the edge capacitance
\be
{\Delta Q(V,T=0) \over V} = - \rho_0 {e^2 \over m}
\label{cap}
\ee
is independent of the sign of $V$ and we establish
the correct value of the Hall conductance
\be
G= {1 \over \rho_0 h} {|\Delta Q| \over V}
={1 \over m} {e^2 \over h}. 
\label{hall}
\ee

To show that the results (\ref{cap}), (\ref{hall}) hold for finite 
temperatures as well we write the general expression
\be
\Delta Q(V,T)=  
- e \rho_0 \int_0^\infty d\eps  \, n_m(\eps + e V) + {e \over m} \rho_0 \int_0^\infty d\eps \, n_{{1 \over m}}(\eps-{e \over m}V)
\ee
and evaluate $\del_{\beta} \Delta Q(V,T)$.
Using once again the duality relations (\ref{dual}), we derive
\be
\del_\beta\Delta Q
= {e \over m} \, \rho_0 \del_{\beta} \int_{-\infty}^\infty
    d\eps \, n_{{1 \over m}} (\eps)
\propto \int_{0}^\infty dx \, \ln(x) \del_x n_{\mm}(x)
\ee
with $x=e^{-\beta\eps}$. Using (\ref{xdelx}) the last 
line turns into
\be
\propto \lim_{x_0 \rightarrow \infty}
 \left[ \ln\la^+(x) - n_{\mm}(x) \ln(x) \right]_0^{x_0}
\ee
and by using the asymptotic behavior for $x\to \infty$
\be
\la^+(x) \approx x^m \ , 
\qquad 
n_{\mm} \approx m
\ee
we conclude that $\del_\beta\Delta Q$ is indeed zero.

%\newpage

\section{Jack polynomials and beyond} 
\setcounter{equation}{0}
\setcounter{figure}{0}

The quasi-particle basis that we specified in (\ref{listofstates})
has some arbitrariness to it. For example, we could have chosen
to act first with the $G_{-t}$ and then with $\phi_{-s}$, which would have 
lead to a different set of states. Also, one quickly finds that the
states (\ref{listofstates}) as they stand are not mutually orthogonal.
For the purpose of establishing the thermodynamics of the fqHe 
edge theory, what matters is the counting of the number of states
with given charge and energy, and this information can be extracted
from (\ref{listofstates}). However, for the analysis of more detailed
questions, in particular those concerning transport, the precise form
of the multi-quasi-particle states is of crucial importance.

In this section, we shall present an `improved' set
of multi-particle states, which are mutually orthogonal and which 
are faithful to the statistics properties of the quasi-particles
$\phi_{-s}$ and $G_{-t}$. The idea will be to specify an operator
$H_{CS}$ that acts on the CFT spectrum, and to modify the multi-particle 
states in such a way that they become eigenstates of $H_{CS}$. The 
operator $H_{CS}$, which was first given by Iso in \cite{Is}, will 
be nothing else than a CFT version of the hamiltonian of so-called 
Calogero-Sutherland (CS) quantum mechanics with inverse square exchange. 
The analogy with CS quantum mechanics confirms the assignment 
of $g={1 \over m}$ ($g=m$) exclusion statistics to $\phi_{-s}$ and 
$G_{-t}$, which are the CFT analogues of the particles and holes of 
the CS system. It also links the Jack polynomial eigenstates of 
the CS system to the quasi-particle basis of the fqHe edge theory. 

We would like to stress that, in the context of the $\nu={1 \over m}$
qHe edge, we do not assign physical significance to the operator 
$H_{CS}$. We merely use this operator as a device to select an
optimal set of multi-particle states, where `optimal' is meant
in the sense of mutual orthogonality and of a relatively simple
form of matrix elements of physical operators between the states.

The need for improving the form of the multi-particle
states (\ref{listofstates}) can be phrased in yet another way.
Let us, as an example, consider a multi-particle state containing
two $\phi$-quanta. If we were to work in position space,
putting the two $\phi$-fields at positions $z_2$ and $z_1$,
the {\em exchange statistics}\ properties of the field $\phi(z)$
would result in simple phase factors associated to the 
interchange $z_2 \leftrightarrow z_1$ in a correlator. 
Working instead in energy space, with $\phi$ quanta $\phi_{-s_2}$ 
and $\phi_{-s_1}$, we expect that the {\em exclusion statistics}\ 
properties of $\phi$ will imply simple behavior under the 
interchange $s_2 \leftrightarrow s_1$ in a correlator or form factor.
In particular, one expects that interchanging $s_2 \leftrightarrow s_1$
in a form factor involving a state $|\ldots,s_2,s_1,\ldots \rangle$ 
will result in a phase factor $e^{i{\pi \over m}}$. We shall show 
below (see, e.g., eq.~\ref{formfac}), that the form factors of the 
true `Jack polynomial' multi-particle states 
$|\ldots,s_2,s_1,\ldots\rangle$ indeed satisfy this simple  
property, which is not valid for the naive multi-particle state 
$(\ldots \phi_{-s_2} \phi_{-s_1}\ldots)|0\rangle$.
In mathematical terms, the issue is to define the
correct co-product in a situation where, due to fractional
statistics, the relevant symmetry is not a Lie algebra but rather 
a quantum group. In the context of the spinon basis for the $m=2$ 
theory, this quantum group is a so-called yangian, and it has been 
established that the `Jack-polynomial' co-product agrees with the 
co-product that is dictated by the quantum group symmetry 
\cite{BPS,BLS}.

\subsection{The operator $H_{CS}$}

To specify the operator $H_{CS}$, we employ the free boson 
$\varphi(z)$, which already featured in our formula (\ref{qp}).
Following \cite{Is}, we define
%\widetext
\be 
   H_{CS} = 
      {m-1 \over m} \sum_{l=0}^\infty (l+1) (i\sqrt{m} 
                \del \varphi)_{-l-1} (i\sqrt{m}\del \varphi)_{l+1}
    + {1 \over 3m} \left[(i\sqrt{m} \del \varphi)^3\right]_0 \ ,
\label{Hcs}
\ee 
%\narrowtext
where $\del \varphi(z) = \sum_l (\del \varphi)_l z^{-l-1}$
and where the second term on the r.h.s. denotes the zero-mode
of the normal ordered product of three factors
$(i \sqrt{m} \del \varphi)(z)$. As a first result, one
finds the following action of $H_{CS}$ on states 
containing a single quasi-particle of charge ${e \over m}$ 
or $-e$
\bea
H_{CS}\, \phi_{-{1 \over 2m}-n} \vac = 
  h_{\phi}(n) \, \phi_{-{1 \over 2m}-n} \vac 
  \ , 
\qquad &&{\rm with}\qquad  h_{\phi}(n) = \left[ {1 \over 3m} + mn (n+{1 \over m})\right]  
\nonu
&&{\rm and}\nonu
H_{CS}\, G_{-{m \over 2}-n} \vac = 
  h_G(n) G_{-{m \over 2}-n} \vac
  \ , 
\qquad &&{\rm with}\qquad  h_G(n) = \left[ - {m^2 \over 3} - n (n+m) \right] \ .
\eea
We would like to stress that the fact that both $\phi_{s}$
and $G_{t}$ diagonalize $H_{CS}$ is quite non-trivial. 
If one evaluates $H_{CS}$ on any vertex operator $\phi^{(Q)}$
(of charge $Q {e \over m}$), one typically runs into the field 
product $(T\phi^{(Q)})(z)$, where $T(z)= - \half (\del \varphi)^2(z)$ 
is the stress-energy of the scalar field $\varphi$. Only for 
$Q=1$ and $Q=-m$ do such terms cancel and do we find that the 
quasi-particle states are eigenstates of $H_{CS}$.

We can now continue and construct eigenstates of $H_{CS}$
which contain several $\phi$ or $G$-quanta. What one
then finds is that the simple product states such as 
(\ref{phistates}) are not $H_{CS}$ eigenstates, but that they 
rather act as head states that need to be supplemented by 
a tail of subleading terms. As an example, one finds 
two-$\phi$ eigenstates to be of the form \cite{BLS}
\be 
|n_2,n_1\rangle =
\phi_{-{3 \over 2m}-n_2}\phi_{-{1 \over 2m}-n_1}|0\rangle 
+ \ \sum_{l=1}^\infty a_l
\left[ \phi_{-{3 \over 2m}-n_2-l}
       \phi_{-{1 \over 2m}-n_1+l} |0\rangle \right]
\ee
with coefficients $a_l$ that can be computed. The connection
of the coefficients $a_l$ with the Jack polynomials that feature in 
the eigenfunctions in CS quantum mechanics has been made 
explicit in \cite{BLS}. For the $H_{CS}$ eigenstate headed by 
the multi-particle state (\ref{listofstates}) (with unit 
coefficient), we shall use the notation
\be  
| \{ m_j\}, \{n_i\} \rangle
\label{jackstates}
\ee
so that
%\widetext
\be
H_{CS} | \{ m_j\}, \{n_i\} \rangle
=
\left[ \sum_{j=1}^M h_G((j-1)m+m_j) 
 + \sum_{i=1}^N h_{\phi}({1 \over m}(i-1)+n_i) \right]
 | \{ m_j\}, \{n_i\} \rangle \ .
\ee
%\narrowtext
Clearly, the states (\ref{jackstates}), with the $m_j$ and $n_i$
as specified in (\ref{listofstates}), form a complete and orthogonal 
basis for the chiral Hilbert space.

\subsection{Norms and form factors}

Of importance for later calculations are the norms of the
states (\ref{jackstates}) and the matrix elements of physical
operators between these states. For the explicit evaluation 
of such quantities we used the connection with Jack polynomials,
relying on results that are available in the mathematical 
literature (\cite{St}, see also \cite{LPS}).

As an example, we focus on multi quasi-hole states 
$| \{ n_i\}\rangle$. To make contact with the Jack's, we
view the ordered set $\{n_i\}$ as a Young tableau $\lambda$. 
The norm-squared of the state $| \{ n_i\}\rangle$ then
becomes
\be
\langle \{n_i\} | \{ n_i\} \rangle = j_{\lambda'}
\label{jlambda}
\ee
where $\lambda'$ is the Young tableau dual to $\lambda$
and the $j_{\lambda}$ are taken from \cite{St}. 
Explicit examples are
%\widetext
\bea
\langle n_1| n_1 \rangle 
&=& j_{(1^{n_1})}
= { (n_1+\mm-1) (n_1+\mm-2) \ldots \mm \over
     n_1 (n_1-1) \ldots 1 }
\nonu
\langle  n_2,n_1  |  n_2,n_1  \rangle 
   &=& j_{(2^{n_1},1^{n_2-n_1})}
\nonu
   &=& { (n_2+\mmm-1)(n_2+\mmm-2) \ldots (n_2+\mmm-n_1) \over
         (n_2+\mm)(n_2+\mm-1) \ldots (n_2+\mm-n_1+1)} 
\nonu
&&   
\times { (n_2-n_1+\mm-1) \ldots \mm \over
         (n_2-n_1) \ldots 1} \, 
       { (n_1+\mm-1) \ldots \mm \over
          n_1 \ldots 1 }
\nonu
{\rm etc.} &&
\eea
In the limit where all $n_i \gg 1$, one finds
\be
\langle \{n_i\}|\{n_i\}\rangle \approx
\prod_{i=1}^N { {n_i}^{\mm-1}  \over \Gamma(\mm)} \ .
\label{jasymp}
\ee

Of interest for the analysis of processes where electrons
or holes tunnel into a fqHe edge is the form factor
\be
  {}_N\langle \{ n_m, \ldots , n_2 , n_1 \} |
  \, G^{\dagger}_{-{m \over 2}-n} \, | \, 0 \, \rangle
  = f(n_m,\ldots,n_1) \, \delta_{n,n_m+\ldots+n_1} \ .
\label{ffdef}
\ee
%\narrowtext
where the subscript $N$ indicates that the state has been 
properly normalized. This form factor describes the 
amplitude by which an incoming hole (described by the operator 
$G^{\dagger}$ and of charge $+e$) creates a state that has 
$m$ quasi-holes excited over the ground state. Explicit 
computation in the limit where all $n_i \gg 1$ yields 
(for simplicity we give the result for $m=3$, see appendix
B for the general case) 
\be
 f(n_3,n_2,n_1) \approx
 {\Gamma({1 \over 3})^{\half} \over \Gamma({2 \over 3})}
 \left( { (n_3-n_2)(n_3-n_1)(n_2-n_1) \over
    n_3 \, n_2  \, n_1 } \right)^{\thi} \ .
\label{formfac}
\ee
Remarkably, this result takes the form of a `Jastrow factor'
in the energy variables $n_i$. The order-(${1 \over 3}$)
zero's when two $n_i$ come near reflect the $g={1 \over 3}$
exclusion statistics properties of the fundamental 
quasi-holes. Note that the expression (\ref{formfac}) is 
invariant under global scalings of all energies $n_i$.
The form (\ref{formfac}) of the form factor can be
viewed as a limit in (chiral) CFT of a result on correlation
functions for the `classical' model of quantum mechanics
with inverse square exchange. This result was conjectured 
by Haldane \cite{Ha3} and later proven in \cite{Ha,LPS}.

%\newpage

\section{Transport properties}
\setcounter{equation}{0}
\setcounter{figure}{0}

Having checked that the thermodynamics of fqHe edges is correctly 
reproduced in the new quasi-particle language we are now ready to move on 
and consider transport properties. Following the set-up of a number 
of recent experiments, we shall consider a situation where electrons (or
holes) from a fermi-liquid reservoir are allowed to tunnel into a $\nu=\mm$
fqHe edge. The DC $I$-$V$ characteristic for this set-up, which were first
computed in by Kane and Fisher \cite{KF} (see also \cite{We2}), show a 
cross-over from a linear (thermal) regime into 
a power-law behavior at high voltages and thus presents a clear fingerprint 
of the Luttinger liquid features of the fqHe edge. The experimental results 
from \cite{CPW} are in agreement with these predictions. (See \cite{dCF} 
for a further theoretical analysis of these data.)

The calculations by Kane and Fisher were based on bosonization and on the 
Keldysh formalism for non-equilibrium transport. Our goal here is to see if 
we can reproduce their results in an approach directly based on the edge 
quasi-particle formalism. Before going into this, we would like to stress 
that the `Thermodynamic Bethe Ansatz (TBA) quasi-particles' behind the 
approach of \cite{dCF} are quite different from what we have here, the 
most important distinction being that the TBA quasi-particles are a 
combination of degrees of freedom of both sides of the tunneling barrier; 
they do not exist for a $\nu=\mm$ edge in isolation.

If the $\nu=\mm$ fqHe edge were to behave as a fermi-liquid, we could 
calculate charge transport across a barrier using a simple (Boltzmann) 
kinetic equation of the form 
%\widetext
\be
I(V,T)\propto e \int_{-\infty}^\infty d\eps \, W \left\{ 
 f_1(\epsilon-eV) F_2(\epsilon) -F_1(\epsilon-eV)f_2(\epsilon) \right\} \ ,
\ee
%\narrowtext
with $f(\eps)$ and $F(\eps)$ the Fermi-Dirac distributions for 
electrons and holes, respectively, and $W$ the probability for 
an electron or hole of energy $\eps$ to cross the barrier and enter 
the edge. As is well known, this Boltzmann equation leads to an 
ohmic (linear in $V$) and temperature-independent current.
Now that we have seen that the non-fermi liquid features of
the $\mm$ edge can be captured via the statistics of the edge 
quasi-particles we can try to write a `Boltzmann equation' for 
transport to and from fqHe edges by putting in appropriate 
generalizations $h(\eps)$ and $H(\eps)$ of the quantities 
$f_2(\eps)$ and $F_2(\eps)$, respectively. Before giving 
precise results (in section 6.1 below) we shall consider a 
`naive' expression based on the intuition from the 
quasi-particle approach. In first approximation, the factor 
$h(\eps)$, which describes the probability for an electron to 
{\em leave}\ a $\nu=\mm$ edge, comprises two effects
\begin{enumerate}
\item
a correlation effect, which can be traced to the non-trivial 
scaling dimension of the edge electron operator (see for example
eq. (\ref{exc})). At zero temperature, this is the so-called 
tunneling density of states
\be
A^+(\eps) \propto \eps^{m-1} \ ,
\ee 
\item
a temperature dependence related to the exclusion 
statistics properties of the edge electrons. As we have
seen, the natural factor associated to the {\em presence}\ 
of an edge electron is the distribution function
\be
n_{g=m}(\eps) \ .
\ee
\end{enumerate}
Combining these factors, we come to the naive expressions
\be
h^{(0)}(\eps) = \eps^{m-1} \, n_{g=m}(\eps) \ ,
\ee
and by similar reasoning we obtain
\be
H^{(0)}(\eps) = \eps^{m-1} \, e^{\beta\eps} \, n_{g=m}(\eps) \ ,
\ee
where the thermal factor $e^{\beta\eps}\, n_m(\eps)$ has been
dictated by the requirement of detailed balance.
\cite{ftn5}

One quickly finds that the Boltzmann equation with factors
$h^{(0)}$ and $H^{(0)}$ is not exact at finite temperature. 
In section 6.2 we shall further comment on this equation 
and argue that it can be viewed as a first stage in a 
systematic approach. Before we come to that, we shall in 
the next section present a particularly simple derivation of 
the exact perturbative $I$-$V$ characteristics for tunneling from a 
Fermi-liquid to a $\nu=\thi$ fqHe edge. This derivation uses 
the idea of a kinetic equation, together with the algebraic 
properties of the edge electrons.

\subsection{Kinetic equation for inter edge transport}

A careful derivation, based directly on the form of the tunneling 
hamiltonian \cite{ftn6}
\be
H_{int} \propto t \, \int d\eps \, 
\left[ \Psi_{\nu=1}^{\dagger}(\eps) \Psi_{\nu={1\over 3}}(\epsilon) 
       + {\rm h.c.} \right] \ ,
\ee
leads to the following kinetic equation (see e.g. \cite{We2})
\be
I(V,T) \propto e \, t^2 \int_{-\infty}^\infty d\epsilon
\left[ f(\epsilon-eV)H(\epsilon)-F(\epsilon-eV)h(\epsilon) \right] \ ,
\label{current}
\ee
where $h,H$ are one particle Green's functions 
\be
H(\epsilon)=\langle\Psi_{\nu={1\over 3}}^\dagger(\eps)
\Psi_{\nu={1\over 3}}(\eps) \rangle_{V,T} \ ,
\qquad
h(\epsilon)=\langle \Psi_{\nu={1\over 3}}(\eps)
\Psi_{\nu={1\over 3}}^\dagger(\eps) \rangle_{V,T}
\label{green1}
\ee
for edge electrons in the $\nu={1\over 3}$ fqHe edge,
taken at $V=0$. Note that the expression (\ref{current})
is perturbative as it gives the lowest non-trivial 
order in the parameter $t$.

The quantities $H(\eps)$ and $h(\eps)$ can be determined by using
two simple observations. The first is that of {\em detailed balance},
which can be phrased as the requirement that at zero voltage there should 
be no current flowing. This fixes the ratio of $H(\eps)$ and $h(\eps)$ 
according to 
\be
H(\epsilon)=e^{\beta(\epsilon-eV)}h(\epsilon) \ .
\ee     
The second observation uses the {\em algebraic properties}\ of the 
edge electron operator, which include the  anti-commutation
relation
%\widetext
\be
\left\{  \Psi_{\nu={1\over 3}}^\dagger(\eps)
,\Psi_{\nu={1\over 3}}(\eps^{\prime}) \right\} 
= {2\pi \over L} {1 \over \rho_0} \epsilon^2 \delta(\eps-\eps^{\prime}) 
  + 6 {E \over \rho_0} + 3 (\eps+\eps^{\prime}) {\Delta Q \over e \rho_0} \ .
\label{N2alg}
\ee
%\narrowtext
In this formula, $E$ is the operator for the total energy per unit 
length (proportional to the Virasoro zero mode $L_0$), and $\Delta Q$ is 
the operator for the total charge per unit length (proportional to
the zero mode $J_0$ of the $U(1)$ Kac-Moody algebra). Clearly, this 
anti-commutator fixes the sum $H(\eps)+h(\eps)$. 
The expectation values of energy and charge follow directly
from our analysis in section 4. We find
\be
\bra E \ket = \rho_0 \left( {\pi^2 \over 6 \beta^2} + {(eV)^2 \over 6} 
                     \right) 
\ , \qquad
\bra \Delta Q \ket = - e \rho_0 {(eV) \over 3} 
\ee
and obtain the exact expressions
\be
H(\epsilon)={ (\epsilon-eV)^2+{\pi^2\over \beta^2} \over 
                       e^{-\beta(\epsilon-eV)}+1 }
\ , \quad
h(\epsilon)={ (\epsilon-eV)^2+{\pi^2\over \beta^2} \over 
                      1+e^{\beta(\epsilon-eV)} } \  .
\label{green2}
\ee
They lead to $I$-$V$ characteristics
\be 
I(V,T)\propto e \, t^2 \, \beta^{-3} \left({\beta eV\over 2\pi}+
\left({\beta eV\over 2\pi}\right)^3\right) \ ,
\ee
in agreement with the result obtained in different approaches 
\cite{KF,dCF}. 

Clearly, the Green's functions (\ref{green1}) can be evaluated in 
other ways, for example by using a conformal transformation in the
$x,t$ domain \cite{We2}. We would like to stress that our derivation 
is more direct and uses nothing more than the fundamental 
anti-commutation relation of the edge electrons.
For $\nu=\thi$, these are particularly simple as they derive from the 
so-called $N=2$ superconformal algebra, which has been well-studied in 
other contexts.
For other filling fractions the fundamental anti-commutators look more 
complicated but are available in principle. 

\subsection{Interpretation in terms of exclusion statistics}

If we compare the exact kinetic equation for 
$\nu=\thi$ with a naive
generalized Boltzmann equation, we see that the mistake in the
latter is in the approximation of the Green's function $h(\eps)$
by a the product $h^{(0)}(\eps)$ of a tunneling density 
of states times a Haldane distribution for fractional statistics.

The reason why this approximation turns out to be rather poor is
that the operator $N(\eps) = \Psi_{\nu=\thi}^\dagger(\eps) 
\Psi_{\nu=\thi}(\eps)$ 
inside a fqHe edge is {\em not}\ to be viewed as a simple counting 
operator weighted by the appropriate power law of $\eps$. This fact 
can be traced to the non-trivial operator terms (proportional to the
energy and the charge operators) in the r.h.s. of (\ref{N2alg}).
To further illustrate this point we evaluated the expectation
value of the operator $N(\eps)$ in a (normalized) one-electron
state $|\eps^\prime\rangle$
\be
\langle \eps^\prime | N(\eps) | \eps^\prime \rangle
 \propto  \eps^2 \delta(\eps-\eps^{\prime}) +
  6 {(\eps^\prime-\eps)(\eps^{\prime \, 2}+\eps^2) \over 
  \eps^{\prime 2}} \theta(\eps^\prime-\eps) \ .
\label{expt}
\ee
This result shows an interaction effect in the action of
$N(\eps)$ on a one-electron state: rather than just counting
quanta of energy $\eps$, the operator $N(\eps)$ is sensitive 
to the presence of quanta at energy $\eps'>\eps$ as well.
In the Green's function $h(\eps)$ (for $\eps>0$), the first 
term on the r.h.s. of (\ref{expt}) corresponds to $h^{(0)}(\eps)$, 
while the second term leads to the following correction term
\be
h^{(1)}(\eps) =
6 \int_{\eps}^{\infty} d\eps^\prime \,
{(\eps^{\prime}-\eps)(\eps^{\prime \, 2}+\eps^2) \over 
\eps^{\prime 2}} \, n_3(\eps^{\prime}) \ .
\ee
In figure~6.1 we have plotted the exact result for 
$h(\eps)$ against
the approximations $h^{(0)}(\eps)$ and $\left[ h^{(0)}+h^{(1)} 
\right](\eps)$. Clearly, the correction term $h^{(1)}(\eps)$ 
greatly improves the accuracy of the description.

\begin{figure}[h]
\setlength{\unitlength}{1pt}
%\centerline{}
%\begin{picture}(200,190)
%\put(-35,-30){\includegraphics{revgreens3.eps}}
%\put(130,10){$\eps \ [\beta^{-1}]$}
\begin{picture}(450,225)
\put(0,-40){\psfig{file=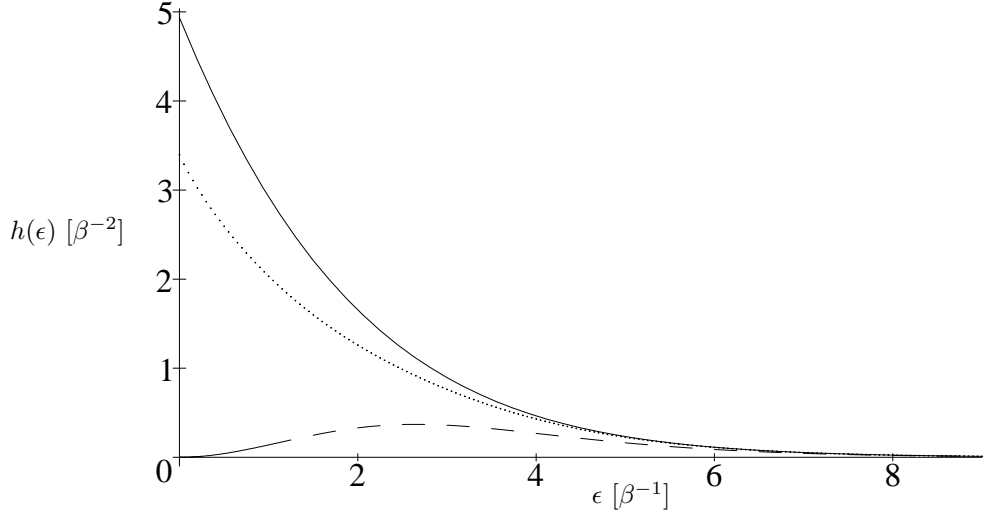}}
\put(230,10){$\eps \ [\beta^{-1}]$}
\put(10,110){$h(\eps) \ [\beta^{-2}] $ }
%\put(-5,85){\rotatebox{90}{ $h( \eps ) \left[\beta^{-2}\right]$}} 
\end{picture}
\caption{One-particle Green's function $h(\eps)$ for filling fraction
   $\nu=\thi$ and at zero voltage. The drawn curve is the exact result
   (\protect\ref{green2}); the dashed curve is the approximation
   $h^{(0)}(\eps)$and the dotted curve corresponds to $\protect\left[
   h^{(0)}+h^{(1)} \protect\right](\eps)$.}
\end{figure}

%\begin{figure}
%\label{hfig}
%\vskip 6cm
%\caption{One-particle Green's function $h(\eps)$ for filling fraction
%$\nu=\thi$ and at zero voltage. The drawn curve is the exact result
%(\ref{green2}); the dashed curve is the approximation $h^{(0)}(\eps)$
%and the dotted curve corresponds to $\left[ h^{(0)}+h^{(1)}\right](\eps)$.
%\end{figure}

The situation here can be described as follows. As
far as thermodynamics goes, the distribution functions 
$n_3(\eps)$ and $n_{\thi}(\eps)$ give exact results for 
quantities such as specific heat and conductance, and we may 
view the edge system as an ideal gas of fractional statistics 
quasi-particles. However, the operators 
$\Psi_{\nu=\thi}^{\dagger}(\eps)$, $\Psi_{\nu=\thi}(\eps)$ 
are not one-particle operators in the usual sense, as they do not 
simply add or extract a single quasi-particle from
a many-particle state. In edge tunneling experiments, the 
edge system communicates with a Fermi liquid {\em via}\ the
operators $\Psi_{\nu=\thi}^\dagger(\eps)$ and 
$\Psi_{\nu=\thi}(\eps)$ and we 
can not avoid interaction effects. We do believe, however, that a 
systematic expansion based on the quasi-particle picture is 
possible.

To avoid reference to a `filled sea of negative energy edge 
electrons' we prefer to discuss transport in the picture where
the fundamental quasi-particles are positive energy edge electrons
and edge quasi-holes, respectively. If we we stick for a
moment to the abovementioned `naive' $0$-th order approximation,
we would arrive at the following lowest contribution to
the tunneling current at voltage $V$ and temperature $T$ 
%\widetext 
\bea
\lefteqn{I^{(0)}(V,T) \propto }
\nonu
&&
- e \int_0^{\infty} d\eps \,
  n_1(\eps-V) \, N_3(\eps) \, \epsilon^2
\nonu
&&
+ e \int_0^{\infty} d \eps \,  
  N_1(\eps-V) \, n_3(\eps) \, \epsilon^2
\nonu
&&
+ e \int_0^{\infty} d\eps_3 \, d\eps_2 \, d\eps_1 \,
   n_1(\sum_i\eps_i +V) \, N_{1 \over 3}(\eps_3) N_{1 \over 3}(\eps_2) 
   N_{1 \over 3}(\eps_1) f^2(\eps_3,\eps_2,\eps_1)
\nonu
&&
- e \int_0^{\infty} d \eps_3 \, d\eps_2 \, d\eps_1 \,
  N_1(\sum_i\eps_i +V) n_{1 \over 3}(\eps_3) n_{1 \over 3}(\eps_2) 
  n_{1 \over 3}(\eps_1) f^2(\eps_3,\eps_2,\eps_1)
\label{b-eq}
\eea
%\narrowtext
where $f(\eps_3,\eps_2,\eps_1)$ is the form factor given in 
(\ref{formfac}) and the integrations are over ordered sets of 
energies $\eps_3\geq\eps_2\geq\eps_1\geq 0$. We have used the 
notation 
\be
N_g(\eps) = e^{\beta \eps} n_g(\eps) \ .
\ee
This result becomes exact in the limit $T\to 0$, where all 
distributions become step functions and the interaction 
effects disappear. Note that this formula has a clear asymmetry 
between electrons and holes: electrons that come into the
edge settle as edge electrons, while incoming holes `decay'
into a total of three edge quasi-holes, with relative amplitudes
given by the form factor $f(\eps_3,\eps_2,\eps_1)$.
For $T=0$, $V>0$ the expression (\ref{b-eq}) reduces to
a single term
\bea
I^{(0)}(V>0,T=0) &\propto&
  - e \int_{0\leq \eps \leq V} d\eps \, \eps^2 \, \nonu  &\propto& \, V^3 \ ,
\eea
while $T=0$, $V<0$ it reduces to
\bea
I^{(0)}(V<0,T=0) \propto
  e \int_{0\leq \sum_i \eps_i \leq V} d\eps_3 \, d\eps_2 \, d\eps_1 \,
  f^2(\eps_3,\eps_2,\eps_1) \,\nonu  \propto \, V^3 \ ,
\eea
In the latter case, the power law $I\propto V^3$ is a simple 
consequence of the fact that we perform three independent 
integrations $\int d\eps_3 d\eps_2 d\eps_1$ over quasi-hole energies, 
with a form factor $f(\eps_3,\eps_2,\eps_1)$ that is scale-invariant. 

Clearly, the expression (\ref{b-eq}) needs corrections. We believe
that a systematic expansion, along the lines of
the expansion $h(\eps)= h^{(0)}(\eps)+h^{(1)}(\eps)+ \ldots $
that we have demonstrated above, is possible. We plan to demonstrate
this in more detail in a future publication.

\section{Conclusions}
\setcounter{equation}{0}
\setcounter{figure}{0}

The edge electrons that have been central in this paper
are the edge analogues of the composite fermions (CF) used
to describe bulk physics. We have made clear that, while the
exchange statistics of these particles are fermionic, their
exclusion statistics properties are not and are instead
captured by non-trivial distribution functions $n_{m}(\eps)$
that take the place of the familiar Fermi-Dirac distribution.
We have also investigated to what extent a quasi-particle
picture, with edge electrons and edge quasi-holes as the 
fundamental quanta, can be used as a starting point for a 
quantitative analysis of transport. We have used algebraic 
properties of the $\nu=\thi$ edge electrons to derive exact 
results, and we have claimed that in general exclusion 
statistics properties may be used to set up a systematic 
expansion. In our view, these results hold some important 
lessons for other situations where fractional statistics 
quasi-particles have been proposed (spinons in $d=1$ quantum 
spin chains, anyons in $d=2$, etc.).

\section*{Acknowledgements} 
We thank A.W.W.~Ludwig for many insightful comments and collaboration
in the early stages of this project. KS thanks P.~Bouwknegt,
E.~Fradkin, and Y.-S.~Wu for discussions. Part of this work was done
at the 1997 ITP Santa Barbara Workshop on `Quantum Field Theory in Low
Dimensions: from Condensed Matter to Particle Physics '.  This
research has further benefited from NATO Collaborative Research
Grant SA.5-2-05(CRG.951303) and by support from the FOM foundation 
of the Netherlands.

\newpage

\appendix

\section{Composite edges: Jain series}
\setcounter{equation}{0}
\setcounter{figure}{0}

In this appendix we briefly describe a quasi-particle
formulation of the composite edge theories corresponding
to the filling fractions $\nu={n \over np+1}$ of the
Jain series. These edge
theories can be written as a collection of $n$
free bosons, coupled via the topological $K$-matrix
of the effective bulk Chern-Simons theory \cite{We1}. 
In \cite{FK} it was shown that the  
effective low-energy CFT for particles satisfying 
Haldane statistics with $n \times n$ statistical 
matrix $G$ is a $c=n$ CFT with topological matrix 
$K=G^{-1}$.  Inverting the argument we expect  
that the fundamental excitations of the CFT for qHe  
matrix $K$ can be interpreted in terms of pseudo-particles  
satisfying fractional exclusion statistics with matrix 
$G=K^{-1}$. 
 
An alternative and more natural approach to the Jain
series edges would be to first perform a change of basis 
which separates a single charged mode from a set of $n-1$ 
neutral modes \cite{KF2,DMM}. The latter are governed by an 
$\widehat{su(n)_1}$ affine Kac-Moody symmetry, and can be 
treated separately. 
An option is to view them as a set of $n$ free parafermions 
in the sense of Gentile, see \cite{Sc}. The CFT for the 
remaining charged mode is of the type that we described in 
this paper, with $g=\nu$. The entire edge theory is then 
described by a single (charged) $g$-on and a set of 
$\widehat{su(n)}_1$ degrees of freedom.
 
As an example of how the chiral Hilbert space works
out, here is the example of $\nu=2/5$, with
$K$ matrix 
\be
K = \left( \begin{array}{cc} 3 & 2 \\ 2 & 3 \end{array} \right) \ .
\label{kmatrix}
\ee
This theory has two independent $U(1)$ affine Kac-Moody symmetries, 
giving a factor $\left[ \prod_{l=1}^{\infty} (1-q^l) \right]^{-2}$ 
in the partition function. The various charge sectors are 
labeled by pairs of integers $(l_1,l_2)$, the energy 
being given by 
$E(l_1,l_2)={1 \over 10}(3l_1^2-4 l_1 l_2+3l_2^2)$ 
(this is the bilinear form defined by the inverse of the
$K$-matrix (\ref{kmatrix})). Thus
\be 
Z_{\nu=2/5}(q)= \sum_{(l_1,l_2)}
  {q^{E(l_1,l_2)} \over 
   \left[ \prod_{l=1}^{\infty} (1-q^l) \right]^2} .
\ee
Under the rearrangement into $\widehat{su(2)}_1$ times $U(1)$,
the combination $\half(l_2-l_1)$ plays the role of
the $su(2)$ spin, while $l_1+l_2$ is the charge under
the new $U(1)$. The character identity will be
\cite{DMM}
\be
Z_{\nu=2/5}(q) = \chi_{j=0}^{su(2)_1}(q) 
  Z_{\rm even}^{2/5}(q)
  + \chi_{j=\half}^{su(2)_1}(q) Z_{\rm odd}^{2/5}(q) \ ,
\label{Z25}
\ee
where the subscript even (odd) on $Z^{s/r}$ means that
we restrict to the states with total $U(1)$ charge $Q$ even 
or odd. Simple expressions for the $su(2)_1$ characters are
\be
\chi_{j=0}^{su(2)_1}(q) = \sum_{m+n \; {\rm even}}
  {q^{\quart (m+n)^2} \over (q)_m (q)_n} \ , \qquad
\chi_{j=\half}^{su(2)_1}(q) = \sum_{m+n \; {\rm odd}}
  {q^{\quart (m+n)^2} \over (q)_m (q)_n} \ .
\ee

For the general case with $\nu={n \over np+1}$, the charged 
sector is described by a free boson CFT at compactification 
radius $R^2=\nu^{-1}$, which we write as $R^2={r \over s}$. 
The chiral partition sum is
\be
Z^{s/r}(q)= \sum_{Q=-\infty}^{\infty}
  {q^{Q^2/(2rs)} \over \prod_{l=1}^{\infty}(1-q^l)} \ ,
\label{Zrs}
\ee
and restrictions, such as the even/odd in (\ref{Z25}) 
are taken into account by restricting the charge quantum
number $Q$.

Our fundamental charged edge quasi-particles will now
be the primary fields of $U(1)$ charges $+s$ and $-r$; 
we shall write the creation and annihilation modes of
these fields as $\phi_{-t}$ and $G_{-t}$, respectively.
Note that for $s\neq 1$ the operators $G_{-t}$ are not the 
physical edge electrons as the latter can only be written 
by including non-trivial factors from the neutral sector!

In close analogy with our analysis in section 3.4, we can
now establish that the states
%\widetext
\bea
&& 
 G_{-(2M-1){r \over 2s}+{Q \over s}-m_M} \ldots
% G_{-3{r \over 2s}+{Q \over s}-m_2} 
 G_{-{r \over 2s}+{Q \over s}-m_1} 
 \phi_{-(2N-1){s \over 2r}-{Q \over r}-n_N} \ldots
% \phi_{-3 {s \over 2r}-{Q \over r}-n_2} 
 \phi_{-{s \over 2r}-{Q \over r}-n_1} | \, Q \, \rangle
\nonumber \\[3mm]
&& \quad {\rm with}\ \qquad 
   m_M \geq m_{M-1} \geq \ldots \geq m_1 \geq 0 ,\quad
   n_N \geq n_{N-1} \geq \ldots \geq n_1 
\nonu
&& \quad {\rm and}\ \qquad
   n_1 \geq 0 \quad {\rm if}\ Q \geq 0
\nonu
&& \qquad \qquad \quad n_1 >0 \quad {\rm if}\ Q < 0 \ ,   
\label{listofstatespq}
\eea
with $Q=-(r-s), \ldots, +(s-1)$,
span the chiral Hilbert space (\ref{Zrs}) of the
charged boson. The total energy of the lowest energy
state in the charge sector $Q$ having particle-numbers 
$M$ and $N$ for the quanta of type $G$ and $\phi$, 
respectively, equals 
\be
E(Q;M,N) = {Q^2 \over 2rs} + {r \over 2s}M^2 -{Q \over s}M
  + {s \over 2r}N^2 + \left[{Q \over r}+\delta_{Q<0}\right]N
\ee           
%\narrowtext
and this leads to the following expression for the 
chiral partition sum
\be
Z^{s/r}(q)= \sum_{Q=-(r-s)}^{(s-1)} \sum_{M,N \geq 0}
  { q^{E(Q;M,N)} \over (q)_M (q)_N } \ .
\label{Zrs2}
\ee
The equality of the expressions (\ref{Zrs}) and (\ref{Zrs2})
is a new identity of the Rogers-Ramanujan type 
(see \cite{KKMM,BLS,BLS2} for some similar identities).

In the case $p<0$, the Jain series qHe edge exhibits 
counterflowing edge modes and it has been claimed that a new
disorder-driven fixed point dominates the physics 
\cite{KFP,KF2}. It will be most interesting to analyze this 
scenario in a quasi-particle formulation. 
 
%\newpage

\section{Form factor for general $m$}
\setcounter{equation}{0}
\setcounter{figure}{0}

We briefly explain the exact evaluation of the form 
factor $f(m_m,\ldots,m_1)$ as defined in (\ref{ffdef}). 
Let us consider
the special case $m=2$ first. In that case the 
`hole operator' $G(z)$ has conformal dimension
$1$ and may be identified with one of the currents
of the affine Kac-Moody algebra $\widehat{su(2)}_1$.
By exploiting the OPE
\be
\phi(z) \phi(w) = 
  (z-w)^{+\half} \left[ G(w) + {\cal O}(z-w) \right] 
\ee
one obtains
\be
G(w) = \oint_{C_w} {dz \over 2\pi i}(z-w)^{-{3 \over 2}}
  \phi(z) \phi(w) \ .
\ee
We also have \cite{BPS}
\be
\phi(z)\phi(w) | \, 0 \, \rangle = (z-w)^{\half}
\sum_{n_2,n_1}  P_{n_2,n_1}^{(-\half)}(z,w) | n_2,n_1 \rangle  \ ,
\ee
where $P_{\{n_i\}}^{(-\half)}(z,w)$ are the appropriate
Jack polynomials. Combining the above, we obtain
\be
G(w) | \, 0 \, \rangle =
\sum_{n_2,n_1} P_{n_2,n_1}^{(-\half)}(w,w) | n_2,n_1 \rangle  
\ee
and it follows that
%\widetext
\be
{}_N\langle  n_2 , n_1  |
  \, G^{\dagger}_{-1-n} \, | \, 0 \, \rangle
  = \delta_{n,n_2+n_1} \,
    \left[ j_{(2^{n_1},1^{n_2-n_1})}\right]^{\half} \, 
    P_{n_2,n_1}^{(-\half)}(1,1) \  ,
\ee
%\narrowtext
with $j_{\lambda'}$ as in (\ref{jlambda}).
For general $m$ one obtains a similar result in terms of Jack
polynomials with label $(-\mm)$. Using the explicit result
\cite{St,LPS}
\bea
P_{\{n_i\}}^{(-\mm)}(1,1) 
  = \prod_{i=0}^{m-1} {\Gamma(\mm) \over 
                       \Gamma(1-{i \over m})}
         \prod_{i<j} {\Gamma(n_j-n_i+ {j-i+1 \over m}) \over
                 \Gamma(n_j-n_i+ {j-i \over m})} \ ,\nonu
\eea
together with the result (\ref{jasymp}) for the
$j_{\lambda'}$, we derive the following asymptotic 
form for $n_i>>1$
\be
f(n_m,\ldots,n_1) = {[ \Gamma(\mm) ]^{m/2} \over 
     \prod_{i=0}^{m-1} \Gamma(1-{i \over m})}
    {\prod_{i<j} {(n_j-n_i)^{\mm}} \over
     \prod_{i=1}^m n_i^{(m-1) \over 2m}} \ .
\ee
The simple Jastrow form of this form factor is a clear 
indication that in the limit $n_i\gg 1$ a much simpler 
derivation, along the lines of `bosonization in momentum 
space' should be possible.

\newpage

\frenchspacing 
\baselineskip=16pt 
%\begin{thebibliography}{11} 


\begin{references}
\bibitem[*]{ftn1} {The familiar terminology `composite fermions'
is somewhat unfortunate in this context, as we argue that 
the {\em exclusion}\ statistics properties of these composite 
quasi-particles are {\em not}\ fermionic. Clearly, their 
{\em exchange}\ statistics, which are determined 
`modulo $2\pi$', are fermionic.}
\bibitem[\dagger]{ftn2}{The notation $G$ is inspired by the fact that the
          fundamental anti-commutation relations for the modes $G_{-t}$ 
          at $\nu=1/3$ are those of the so-called $N=2$ superconformal 
          algebra. See section 6 for more on this.}
\bibitem[\S]{ftn4}{We use the symbol $\Delta Q$ for the total charge,
while we keep $Q$ for the reduced charge (in units of ${e \over m}$).}
\bibitem[\P]{ftn5}{This same thermal factor was proposed in the paper
  \cite{BBM}, which proposes a generalization of the Boltzmann
  equation to the case of fractional statistics. In our
  view, the proposal of \cite{BBM} is incomplete, as it ignores 
  the correlation effects which are unavoidable for 
  quasi-particles obeying fractional exclusion statistics.}
\bibitem[\|]{ftn6}{We write $\Psi$ for the edge electron operator,
(denoted by $\Psi_e$ in section 2 and by $G$ in section 3),
and we indicate the filling fraction by an explicit subscript.}
\end{references}
\end{document}